\documentclass[mreferee]{gji}
\usepackage{timet}

\usepackage{graphicx}
\usepackage{amssymb}
\newcommand{\p}{\partial}
\newcommand{\n}{\nabla}
\newcommand{\vpm}{v_+^{(-)}}
\newcommand{\php}{(t+\varphi)}

\title[Geophys.\ J.\ Int.: Visco-magnetic Torque at the CMB]
  {Visco-magnetic torque
 at the core mantle boundary}
\author[B. Deleplace and P. Cardin]
  {B\'erang\`ere Deleplace and Philippe Cardin\\
  Laboratoire de G\'eophysique Interne et Tectonophysique,
Observatoire de Grenoble,\\
Universit\'e Joseph Fourier and CNRS,
Grenoble, France.
  }
\date{Received 1998 December 18; in original form 1998 November 22}
\pagerange{\pageref{firstpage}--\pageref{lastpage}}
\volume{200}
\pubyear{1998}

\begin{document}

\maketitle

\begin{abstract}
 
A magneto-hydrodynamic model of boundary layers at the Core-Mantle Boundary (CMB) is derived and used to compute the viscous and electromagnetic torques generated by the Earth's nutation forcing. The predicted electromagnetic torque alone cannot account for the dissipation estimated from the observations of the free core nutation. The presence of a viscous boundary layer in the electromagnetic skin layer at the CMB, with its additional dissipative torques, may explain the geodetic data. An apparent Ekman number at the top of the core between $2$ and $4\;10^{-11}$ is inferred depending on the electrical conductivity of the mantle.   
 
\end{abstract}

\section{Introduction}

Detailed models of coupling at the Core Mantle Boundary (CMB) have been put forward to explain the more and more accurate measurements of the nutations of the Earth \cite{wahr81,deha97,math02}. The nutations of the Earth induce a differential rotation, about an equatorial axis, between the mantle and the core \cite{sasa80,buff92}. This differential rotation at the CMB generates both a viscous torque \cite{gree68,lope75,roch76} and an electromagnetic torque due to the shear of the poloidal magnetic field lines \cite{roch60,toom74,sasa77}. Buffett and his colleagues developed sophisticated models of the electromagnetic torque at the CMB \cite{buff92,buff93,buff02} in order to fit the spatial geodetic observations. First, Buffett \shortcite{buff92} introduced a weak magnetic field theory  where the Lorentz forces associated to the skin magnetic effect are too small to generate any motion in the boundary layer. His magnetic analysis requires the presence of a very good electrically conducting layer in the lowermost mantle (same electrical conductivity as the core) to get an adequate amplitude of the torque. Moreover, Buffett \shortcite{buff92} invoked a enhanced magnetic field at the CMB (4 times larger than the observed one) to account for the small scales of the magnetic field. The value of the small scales of the magnetic field at the CMB (spherical harmonic degree $l>13$) cannot be measured at the surface of the Earth because the crustal magnetic field is dominant at these wavelengths \cite{blox95,stac92}. He estimated their effect using an extrapolation of the low-degree non dipole part of the poloidal magnetic spectrum to higher degrees. Then, Buffett \shortcite{buff93} investigated the role of a toroidal magnetic field on the electromagnetic torque at the CMB. Its effects are weak and do not increase the dissipation of magnetic origin at the CMB. Moreover, his results are rather speculative as measurements of the toroidal magnetic field in the Earth's core are not available. Buffett et al. \shortcite{buff02} improved the 1992's model by relaxing the weak field approximation. Thus, they solved the inviscid dynamics of the skin layer in the presence of Lorentz forces. The ratio of the velocity induced by the Lorentz forces in the skin layer and the velocity jump at the CMB is of the order of the Elsasser number (defined below). Its value, at the top of the core, is comprised between $0.1$ and $1$ so that the weak field approximation is not valid. The presence of this dynamical effect reduces the amplitude of the electromagnetic torque at the CMB. This is the reason why, in order to fit the improved observational constraints \cite{math02}, Buffett et al. \shortcite{buff02} invoked the presence of a constant magnetic field modeling the non dipole component (small scale magnetic field), three times greater than the dipole value at the CMB. In all their studies, Buffett and his colleagues introduced a thin electrically conducting layer at the base of the mantle. Its presence remains necessary to get the correct amplitude of the electromagnetic torque.

For rapidly rotating fluids, viscosity plays a role mainly in thin boundary layers, the so-called Ekman layers \cite{gree68}. The depth of these layers is $\sqrt{\nu/\Omega}$ where $\nu$ is the kinematic viscosity of the fluid and $\Omega$ the angular velocity of the Earth. As the magnetic skin depth is $\sqrt{\eta/\Omega}$, where $\eta$ is the magnetic diffusivity of the core, the ratio of the two lengths is given by $\sqrt{P_m}$ where $P_m = \nu/\eta$ is the magnetic Prandtl number. Table \ref{tab: phprop} contains the  values of the molecular diffusivities for the core \cite{poir94}. We  evaluate $P_m = 4 \;10^{-6}$ in the core, making the viscous layer 500 times thinner than the magnetic skin layer.

\begin{table}
\begin{tabular}{|l|l|l|}
\hline
&parameter & value \\
\hline
$R$			 	&	core radius														& $3.48 \;10^6 \;m $\\
$\Omega$	& rotation rate of the Earth						& $7.29 \; 10^{-5} \;rad \;s^{-1}$\\
$\rho$		& density 															& $10^4 \;kg \;m^{-3}$\\
$\eta$		& magnetic diffusivity of the core 			& $1.6 \;m^2 \;s^{-1}$\\
$\eta_M$	& magnetic diffusivity of the mantle  	& $1.6-1600 \;m^2 \;s^{-1}$\\
$\nu$			& kinematic viscosity of the core 			& $7.0 \; 10^{-6} \;m^2 \;s^{-1}$\\
$B_0$			& magnetic field at the CMB							& $0.46 \; 10^{-3} \;T$\\
$K^{CMB}$ & coupling constant at the CMB 					& $-1.85 \;10^{-5}$\\ 
$E$				& Ekman number  
                  $\frac{\nu}{\Omega R^2}$				& $8.0 \;10^{-15}$\\
$E_m$			& magnetic Ekman number of the core			
                  $\frac{\eta}{\Omega R^2}$       & $1.8 \;10^{-9}$\\
$E_m^M$		& magnetic Ekman number of the mantle		
      						$\frac{\eta_M}{\Omega R^2}$	    & $1.8 \;10^{-9}-1.8 \;10^{-6}$\\
$\Lambda$	& Elsasser number												
          $\frac{\sigma {B_0}^2}{\rho \Omega}$		& $0.14$\\
$P_m$			&	magnetic Prandtl number								
									$\frac{\nu}{\eta}$							& $4.5 \;10^{-6}$\\
\hline
\end{tabular}
\caption{Physical properties and associated dimensionless numbers used in this study.}
\label{tab: phprop}
\end{table}

Recent numerical simulations of the geodynamo have been successful in reproducing some features of the magnetic field of the Earth \cite{dorm00}. They have in common to use a very high viscosity (Ekman number greater than $10^{-6}$) so as to avoid numerical resolution problems. Glatzmaier \& Roberts \shortcite{glat95} advocated the use of an eddy viscosity for dynamical core modeling as it is generally done in numerical modeling of the oceanic or atmospheric sciences \cite{pedl87}. Brito at al. \shortcite{brit04} have found evidence of apparent viscosity from an experiment of thermal convection in a rapidly rotating spherical shell filled with water using a spin-up technique. They interpret their observations by arguing that turbulent motions in the bulk of the core increase the efficiency of the exchange of angular momentum between the Ekman layers and the geostrophic volume. These non-linear effects at small scale may be modeled by an eddy viscosity at large scale. A turbulent viscosity at the top of the core between $10^{-4} \;m^2 \;s^{-1}$ and $10^{-1} \;m^2 \;s^{-1}$ is possible. Such eddy viscosities increase the magnetic Prandtl number and decreases the ratio between the "viscous" and magnetic layer depths. Under these conditions, viscous effects have to be incorporated in the dynamical equation of the layer. With such a theory, the quality factor of the free core nutation, deduced from the geodetic data, is a constraint on the apparent viscosity et the top of the core for the diurnal frequency.

Recently, two related studies \cite{math05,palm05} have been published. Their approaches are very similar to the work presented here and lead also to the prediction of a viscosity value at the top of the core from nutations data. From their own data analysis, Palmer \& Smylie \shortcite{palm05} use an approximate viscous model to infer a viscosity. Mathews \& Guo \shortcite{math05} introduce a magneto-viscous model similar to ours and determine the viscosity from the observational data analysis of Mathews et al. \shortcite{math02}. Both papers give a value of viscosity which is close to the one proposed in this paper. However, our analysis proposes a complete calculation with all spectral components of the magnetic field with different extrapolated tendencies for the hidden part ($l>13$) of the magnetic field at the CMB. Moreover, a physical description of the magnetic and viscous boundary layers is shown. We also give a complete study of the variations of the electrical conductivity at the base of the mantle which enables us to invert the observational data to obtain trade-offs between the mantle electrical conductivity effect and the viscous effect at the CMB.

This paper presents a derivation of a magneto-hydrodynamic boundary layer attached to the mantle taking into account the Lorentz, Coriolis and viscous forces (section 2). In section 3, we discuss the influence of the geometry (small scales) and amplitude of the magnetic field at the CMB on the electromagnetic torque. Section 4 describes the effects of a viscous layer on the visco-magnetic torques at the CMB and an Ekman number is estimated at the top of the core. The variations of the electrical conductivity in the lowermost mantle are studied in section 5. A final discussion ends the paper.

\section{Mathematical formulation of the torques}

At first order \cite{poin10}, the response of the rotating fluid core to Earth's nutations is a rigid body rotation. This approximation was checked experimentally \cite{vany95} and stays valid for large forcings \cite{noir03}. In the computation of the electromagnetic and viscous torques at the core mantle boundary (CMB), we may neglect the flow induced by the ellipticity of the CMB \cite{sasa80,buff02} and we describe the main flow in the outer core by an angular velocity ${\bf \Omega}$.
We consider the magnetohydrodynamical equations in the frame of reference $({\bf e_x,e_y,e_z})$ rotating with the fluid outer core at the angular velocity vector ${\bf \Omega}$, ${\bf e_z}$ being defined by ${\bf e_z} = {\bf \Omega}/\Omega$. 
The equations are made dimensionless using $\Omega^{-1}$ as time scale, $R$ the radius of the core as length scale and a typical magnitude of the radial component of the magnetic field $B_0$ as magnetic field scale. The  magnetic field and the flow velocity in the core ($r<1$) are governed by the following dimensionless equations:
\begin{eqnarray}	
	\label{B}
	\frac{\p {{\bf B}}}{\p t}+({\bf v} \cdot \n)  {\bf B}&=& ({\bf B}\cdot\n)  {\bf v}  + E_m \Delta \bf{B}\;,
	\\
	\label{v}	
	\frac{D {\bf v}}{D t}+ 2 {\bf e_z} \times {\bf v}
	+ \frac{\p {\bf \Omega}}{\p t}\times {\bf r} + {\bf \Omega}\times\left({\bf \Omega}\times{\bf r}\right) 
	&=&-\n {\bf P} + 
	E_m \Lambda (\n \times {\bf B} )\times {\bf B} + E \Delta {\bf v} \;,\;\;\;\;\;
\end{eqnarray}
where
 $$ E=\frac{\nu}{\Omega R^2}$$
 is the Ekman number and $\nu$ the kinematic viscosity.
$$ E_m=\frac{\eta}{\Omega R^2}= \frac{\eta}{\nu} E $$
is  the magnetic Ekman number which is Ekman number over the magnetic Prandtl number ($\nu/\eta$) where $\eta$ is the magnetic diffusivity.
$$ \Lambda=\frac{\sigma {B_0}^2}{\rho \Omega} $$
is the Elsasser number, $\sigma = (\mu_0 \eta)^{-1}$ is the electrical conductivity of the core and $\rho$ the density of the core fluid.

In the above defined frame of coordinates , the motion of the mantle is a rigid body rotation $\mitbf{\delta\omega}_M$ rotating at $-{\bf e_z}$ defined by :
$$
\mitbf{\delta\omega}_M(t) = \delta\omega_M [{\bf e_x}\cos t - {\bf e_y}\sin t]
$$
The angular velocity of the mantle is equatorial (no spin-up contribution) \cite{buss68,noir03}.
The dimensionless velocity in the mantle is described by:
\begin{equation}
{\bf v_M}=\mitbf{\delta\omega}_M \times {\bf r} = - r \delta\omega_M [{\bf e_\theta}\sin(t+\varphi) + {\bf e_\varphi}\cos\theta \cos(t+\varphi)]
\label{V_M}
\end{equation}
where $({\bf e_r,e_\theta,e_\varphi})$ is the spherical coordinate system directly associated to $({\bf e_x,e_y,e_z})$. 
The magnetic field in the mantle ($r>1$) is then described by the induction equation :
\begin{equation}
\label{B_M}
	\frac{\p {{\bf B}}}{\p t}+({\bf v}_M \cdot \n)  {\bf B}= ({\bf B}\cdot\n)  {\bf v}_M  + E_m^M \Delta \bf{B}
\end{equation}
where 
$$ E_m^M=\frac{\eta_M}{\Omega R^2}$$
is the magnetic Ekman number of the mantle and $\eta_M$ is the magnetic diffusivity of the mantle.

When $\delta\omega_M = 0$, the solution of (\ref{B}) and (\ref{B_M}) is a diffusive poloidal magnetic field denoted ${\bf B}_0$. As $\delta\omega_M$ increases,
 magneto-viscous boundary layers develop around the core mantle boundary ($r=1$). The induced magnetic field in 
 these boundary layers is denoted ${\bf b}$. As $\delta\omega_M << 1$, we have $ b << B_0 = O(1)$. The width of the magnetic skin layer at the top of the core 
 (at the bottom of the mantle) is of order ${E_m}^{1/2}$ ($(E_m^M)^{1/2}$) which is very small compared to 1. The viscous layer of size $E^{1/2}$ is even smaller. Consequently, only radial derivatives of $b$ and $v$ have to be considered in the magneto hydrodynamic equations in the boundary layers. Moreover, we neglect the radial variations of ${\bf B}_0$ and ${\bf v}_M$ inside these thin boundary layers. A linearisation of equations (\ref{B}),(\ref{v}),(\ref{B_M}) with the above boundary layer assumptions leads to: 
\begin{eqnarray}
	\label{inducore}
	\forall r<1, &\qquad& \frac{\p {\bf b}}{\p t}- E_m   \frac{\p^2}{\p r^2} {\bf b} = \beta \frac{\p {\bf v}}  {\p r} 
	\\
	\label{NScore}	
	\forall r<1, && \frac{\p {\bf v}}{\p t}+ 2 {\bf e_z} \times {\bf v} - E\frac{\p^2}{\p r^2} {\bf v}= -\n {\bf \Pi} + E_{m} \Lambda \left(\beta \frac{\p {\bf b}}{\p r} - {\bf B}\cdot \frac{\p {\bf b}}{\p r} \; {\bf e_r}\right)\qquad
	\\	
	\label{indumantle}	
	\forall r>1, &\qquad& \frac{\p {\bf b}}{\p t}- E_m^M \frac{\p^2}{\p r^2} {\bf b} = 0
\end{eqnarray}
where $\beta(\theta,\varphi) = {\bf B}_0(r=1) \cdot {\bf e_r}$ is the radial component of the imposed magnetic field.

Taking the curl of the motion equation (\ref{NScore}) to eliminate pressure and using equation (\ref{inducore}) to eliminate {\bf b}, we obtain the following equation:
$$
 (\frac{\p}{\p t}-E\frac{\p^2}{\p r^2})(\frac{\p}{\p t} -E_m\frac{\p^2}{\p r^2})(\n \times {\bf v})
 -2 \cos\theta (\frac{\p}{\p t}-E_m\frac{\p^2}{\p r^2}) \frac{\p \bf v}{\p r} =  E_m \Lambda \beta^2\frac{\p(\n \times \bf v)}{\p r}
 $$
Under our assumptions, 
$\n \times {\bf v} = -\frac{\p v_\varphi}{\p r}{\bf e}_\theta + \frac{\p v_\theta}{\p r}{\bf e}_\varphi$
and it is convenient to use the complex variables $v_+=v_\theta +iv_\varphi$, and $v_-=v_\theta -iv_\varphi$ to rewrite the last equation:
\begin{equation}
 \label{equa4}
 \frac{\p}{\p r}
 \left[(\frac{\p}{\p t}-E\frac{\p^2}{\p r^2})(\frac{\p}{\p t} -E_m\frac{\p^2}{\p r^2}) v_{\pm} 
 \pm 2i \cos\theta (\frac{\p}{\p t}-E_m\frac{\p^2}{\p r^2}) v_{\pm}\right] 
 =  E_m \Lambda{\beta}^2\frac{\p^3 v_{\pm}}{\p r^3} \;, 	
 \end{equation}
with the imposed velocity (in terms of complex variables) as boundary condition ($r=1$) deduced from equation (\ref{V_M}):
$$
v_{M\pm}=\frac{i\delta\omega_M}{2}[(1 \mp \cos\theta)\exp{i\php} - (1 \pm \cos\theta)\exp{-i\php}] \;.
$$

The boundary condition imposes a time dependence of the form $\exp{\pm it}$ which leads to a set of four differential equations deduced from (\ref{equa4}). We use the exponent $^{(\pm)}$ to denote the sign of the time dependence and we define: 
$$
v_{\pm} = v_{\pm}^{(+)}\exp{it} + v_{\pm}^{(-)}\exp{-it}
$$
In the following, we solve explicitly the problem for one component of the velocity $\vpm$ and the  other three components of the velocity may be easily deduced from it. The equation for $\vpm$ is:
$$
 \frac{\p}{\p r} \left[
 E E_m \frac{\p^4}{\p r^4} +
 (iE +iE_m -2iE_m\cos\theta -E_m \Lambda \beta^2) \frac{\p^2}{\p r^2} 
 + 2 \cos\theta -1 
 \right]\vpm =0 \;.
$$
The associated polynomial function has four roots $Z_i$ corresponding to four exponential elementary solutions which can be written (thanks to M. Greff):
\begin{eqnarray}	
Z_i &=& \pm (2EE_m)^{-1/2} \left[-iE - E_m(i -2i\cos\theta -\Lambda \beta^2)\right. \nonumber \\ \nonumber 
& & \pm \left[-E^2 + 2EE_m(1 - 2\cos\theta -i\Lambda \beta^2)+ \right.\\ \nonumber
& & \left. \left. E_m^2(4\cos\theta -1 -2i\Lambda \beta^2 -4\cos^2\theta + 4i\cos\theta \Lambda \beta^2 + \Lambda^2 \beta^4)\right]^{1/2}\right]^{1/2}
\end{eqnarray}	

As the velocity field must vanish far away from the boundary layer ($\lim_{r E^{-1/2} \rightarrow -\infty} v = 0$), both roots with a positive real value are retained and the solution may be written:
$$
\forall r<1, \qquad \vpm = X_1 \exp(Z_1 (r-1)) + X_2 \exp(Z_2 (r-1)) \;,
$$
where $X_1$and $X_2$ are constants to be determined.

The equation (\ref{inducore}) gives us the solution for $b_+^{(-)}$:
$$
\forall r<1, \qquad b_+^{(-)} = -\frac{\beta Z_1}{i+E_m Z_1^2} X_1 \exp(Z_1 (r-1)) - \frac{\beta Z_2}{i+E_m Z_2^2} X_2 \exp(Z_2 (r-1)) \;. 
$$

Using $\lim_{r(E_m^M)^{-1/2} \rightarrow \infty} b = 0$, the solution for the magnetic field in the mantle may be deduced directly from (\ref{indumantle}):
$$
\forall r>1, \qquad b_+^{(-)} = X_3 \exp(Z_3 (r-1)) \;,
$$
where $Z_3 = -(1+i)/\sqrt{2 E_m^M}$.

We use the continuity of the velocity, the magnetic field and the electrical currents at the core mantle boundary ($r=1$) to determine the constants $X_1, X_2, X_3$.
\begin{eqnarray}	
	X_1  + X_2 &=& \frac{i\delta\omega_M}{2}(1 + \cos\theta) \;,\nonumber\\
	X_3  &=& -\frac{\beta Z_1}{i+E_m Z_1^2} X_1  - \frac{\beta Z_2}{i+E_m Z_2^2} X_2  \;,\nonumber\\
	E_m^M X_3 Z_3  &=& E_m \left[-\frac{\beta Z_1^2}{i+E_m Z_1^2} X_1  - \frac{\beta Z_2^2}{i+E_m Z_2^2} X_2  \right] \;.\nonumber
\end{eqnarray}
 
With the solutions to this set of equations, the velocity and the magnetic field are fully determined within the boundary layers.
With our scaling, the magnetic torque $\Gamma_m$ scales with $\rho R^5 \Omega^2 E_m \Lambda $ and the viscous
torque $\Gamma_v$ with $\rho R^5 \Omega^2 E $.
Here, we use the complex notation by introducing  $\bar \Gamma =\Gamma_x + i\Gamma_y$. The derivation of the viscous torque is given in Appendix A.
\begin{eqnarray}
\bar \Gamma_v &=& ({\bf e_x} + i {\bf e_y}) \cdot \int\int_{S}{\bf r} \times {\bf f}_v d{\bf S}\label{VT} \;,\\
              &=& \frac{i}{2}\int_0^{\pi} \int_0^{2\pi}[(1+\cos\theta) \frac{\p v_+}{\p r}+
(1-\cos\theta) \frac{\p v_-}{\p r}]\exp{(i\varphi)}\sin{\theta}d\theta d\varphi\;.\label{VT2}
\end{eqnarray}

The magnetic torque could be calculated by a surface integral \cite{roch62} similarly to the viscous torque. The magnetic torque could be deduced from the perturbed magnetic field ${\bf b}$ at $r=1$:  
\begin{eqnarray}
\bar \Gamma_m&=& ({\bf e_x} + i {\bf e_y}) \cdot \int {\bf r} \times (\beta {\bf b}) d S  \;, \label{MT}\\ 
              &=&  \frac{i}{2}\int_0^{\pi} \int_0^{2\pi} \beta [(1+\cos\theta)b_+
+(1-\cos\theta)b_-]\exp{(i\varphi)}\sin{\theta}d\theta d\varphi \;.
\label{CEM}
\end{eqnarray}

It is of some use to introduce the coupling constant $K$ deduced from the torque to compare with the observed data \cite{math02}:
$$
K = \frac{\bar \Gamma}{i I \delta\omega_M}  \;,
$$
where $I$ is the dimensionless moment of inertia of the core.

Both torques are integrated numerically using a $(\theta,\varphi)$ grid where $\beta(\theta,\varphi)$ is prescribed. 

For a very weak magnetic field (the Lorentz forces tend to vanish) Toomre \shortcite{toom74} predicted that the torque is pointing $\pi/4$ away from the direction of the imposed angular velocity \mitbf{\delta\omega}. For very low Elsasser and Ekman numbers, the solution follows this asymptotic behavior ($Im(K)=-Re(K)$). For a large and dipolar magnetic field, we compare successfully our results for a very low Ekman number ($E=10^{-16}$) with the coupling constants found by Buffett \shortcite{buff02} in their inviscid study. In the limit of large Ekman numbers and low Elsasser numbers, we check that the computed torque tends toward the spin-over torque \cite{gree68}. 

\section{Magnetic field at the CMB}

\begin{figure}
	\begin{center}
		\includegraphics [width = 12cm] {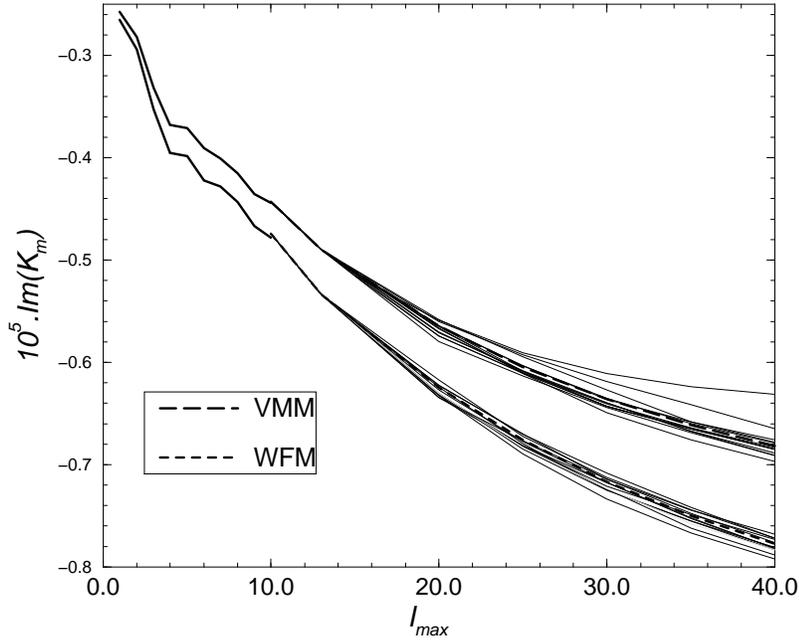}
	\end{center}
	\caption{Imaginary part of the electromagnetic coupling constant versus the truncature level $l_{max}$ of the magnetic field at the core boundary. The observed geomagnetic spectra is extrapolated randomly with a power law $(0.959)^l$ for $l>13$ and the coupling constant is computed with the visco-magnetic model (VMM) or with the weak field model (WFM). Each line represents a set of spherical harmonic coefficients sastifying the spectra dependence. The bold lines show the mean values of both model. }
	\label{fig:B1}
\end{figure}

\begin{figure}
	\begin{center}
		\includegraphics [width = 12cm] {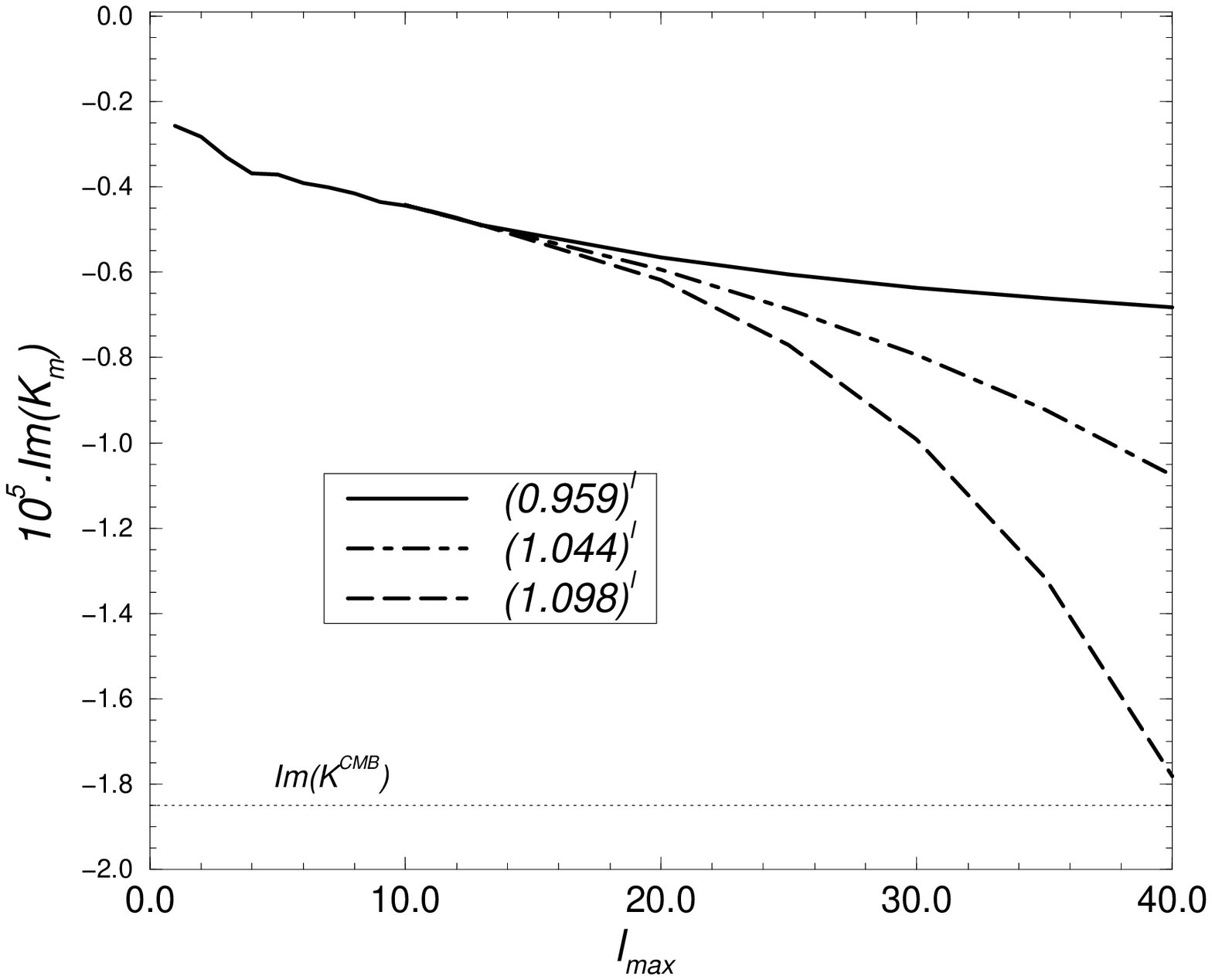}
	\end{center}
	\caption{Imaginary part of a mean electromagnetic coupling constant versus the truncature level $l_{max}$ of the magnetic field at the core boundary. The coupling constant is computed with the visco-magnetic model with three different extrapolated spectra for $l>13$.}
	\label{fig:B2}
\end{figure}

Even though the magnetic field at the CMB is dominated by the axial dipole component, all spherical harmonic components contribute to the electromagnetic torque. The spatial magnetic power spectrum (Mauersberger-Lowes spectrum) of the magnetic field at the Earth surface is deduced from observations \cite{lang82,voor02}. At the CMB, this spectrum is fitted by the power law $1.085\;10^{10}\;(0.959)^l nT^2$ \cite{stac92} if one excepts the dipole component ($l=1$). This is a relatively flat magnetic spectrum which has to become stiffer at very large $l$ to satisfy energetic arguments \cite{robe03,chri04}. Taking the same electrical conductivity for the core and the mantle ($E_m = E_m^M = 1.8 \;10^{-9}$) and a very low Ekman number ($E=10^{-16}$), we compute the electromagnetic coupling constant using the first $l_{max}$ degrees of the magnetic field (higher degree coeficients are set to zero). 
Figure \ref{fig:B1} illustrates possible contributions of the small scales ($l>13$) of the magnetic field to the amplitude of the electromagnetic torque. Different random sets of spherical harmonic coefficients matching the spectra dependence, give comparable contribution to the torque (variations lower than 10\%). Consequently, the mean value of the coupling constant is representative of what could happen at the CMB and in the following, we keep only the mean value to present the results.

The contribution of the large degrees of the spherical harmonics of the magnetic field in the visco-magnetic model is smaller that the one associated with the weak field model \cite{buff92}. The back reaction of the Lorentz forces on the flow is to reduce the electromagnetic torque at the boundary. This effect is emphasised for the small scales of the magnetic field. In some cases (for example, all coefficients positive), the contribution of the large degree is negligeable and the coupling constant curve becomes flat (highest curve in Figure \ref{fig:B1}). 

Figure \ref{fig:B2} shows that the mean electromagnetic torque associated with the observed magnetic field at the CMB ($(0.959)^l$) is too low to fit the imaginary part of the observed coupling constant ($-1.85 \;10^{-5}$). Following the ideas of Buffett \shortcite{buff92,buff02}, we explore the effect of an increase of the magnetic field at small scales . We studied two different power laws for the spectra ($l>13$): $(1.044)^l$ giving an magnetic energy 10 times greater than the standard one at degree 40, and $(1.098)^l$ corresponding to an energy 10 times greater than the standard one at degree 30. An increase by a factor 10 in energy at degree 40, increases the overall coupling constant less than a factor 2 which remains too small to match the observed value. In Figure \ref{fig:B2}, we see that only the $(1.098)^l$ spectrum could explain the observed data. From a geophysical point of view, this spectrum is unlikely as it dissipates a large amount of energy. Using the result of Roberts et al. \shortcite{robe03} (eq 2.7 page 104), we found a ohmic dissipation of $0.03 TW$ for $l<40$. This is large compared to the dissipation associated to dipolar component alone which is $0.08 GW$. According to the scaling deduced from numerical dynamos \cite{buff02b,robe03,chri04} which takes into account the dissipation of the toroidal part of the magnetic field and the contribution of small scales of the magnetic field, this $(1.098)^l$ spectrum is too dissipative. For example, with a dissipation of $1.32 GW$ associated to the large scale magnetic spectrum, Roberts at al. \shortcite{robe03} estimate a total ohmic power loss between 1 and $2TW$.\\
Another dissipative process is thus needed to explain the nutation data.

\section{Viscous effects at the CMB}

\begin{figure}
	\begin{center}
		\includegraphics [width = 12cm] {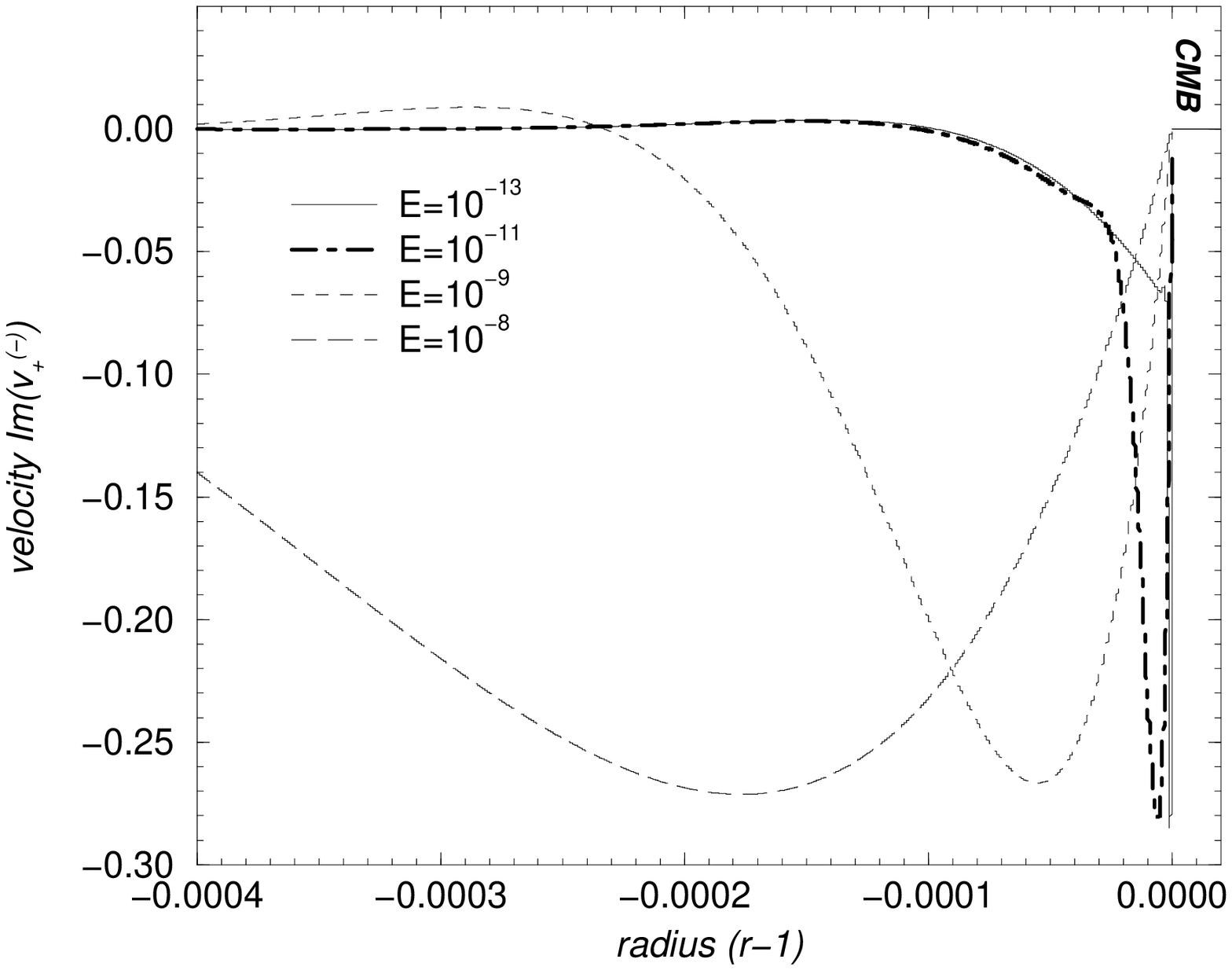}
		\includegraphics [width = 12cm] {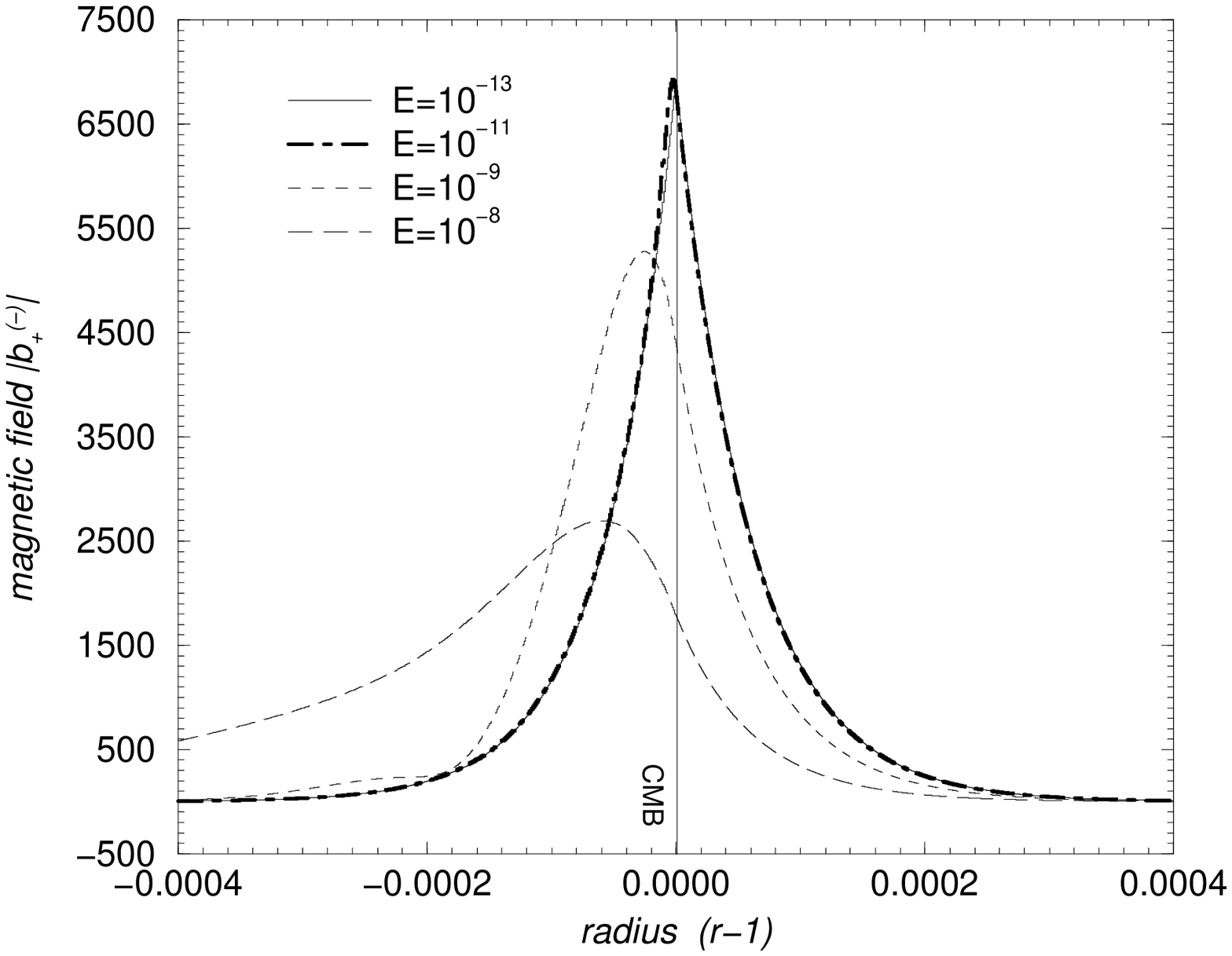}
	\end{center}
	\caption{Velocity $Im(\vpm)$ and magnetic $|b_+^{(-)}|$ perturbed fields at the core-mantle boundary for $E_m = E_m^M = 1.8 \;10^{-9}$ and different Ekman numbers.}
	\label{fig:couche}
\end{figure}

In this section, we assume the electrical conductivity in the core and in the mantle to be the same \cite{buff02} and focus on viscous effects. Figure \ref{fig:couche} shows components of the velocity and perturbed magnetic fields in the boundary layers for $E_m = E_m^M = 1.8 \;10^{-9}$ (see table \ref{tab: phprop}) and different Ekman numbers. For very low Ekman numbers, the viscous layer is very narrow ($E^{1/2}$) and the magnetic field (symmetry and amplitude) is nearly unchanged by the presence of the Ekman layer. For Ekman numbers comparable to the magnetic Ekman number, the width of the viscous layer becomes as large as the magnetic skin depth ($E_m^{1/2}$) and magnetic field is induced deeper into the core. Consequently, the perturbed magnetic field looses its symmetry and its value at the CMB decreases. 

\begin{figure}
	\begin{center}
		\includegraphics [width = 12cm] {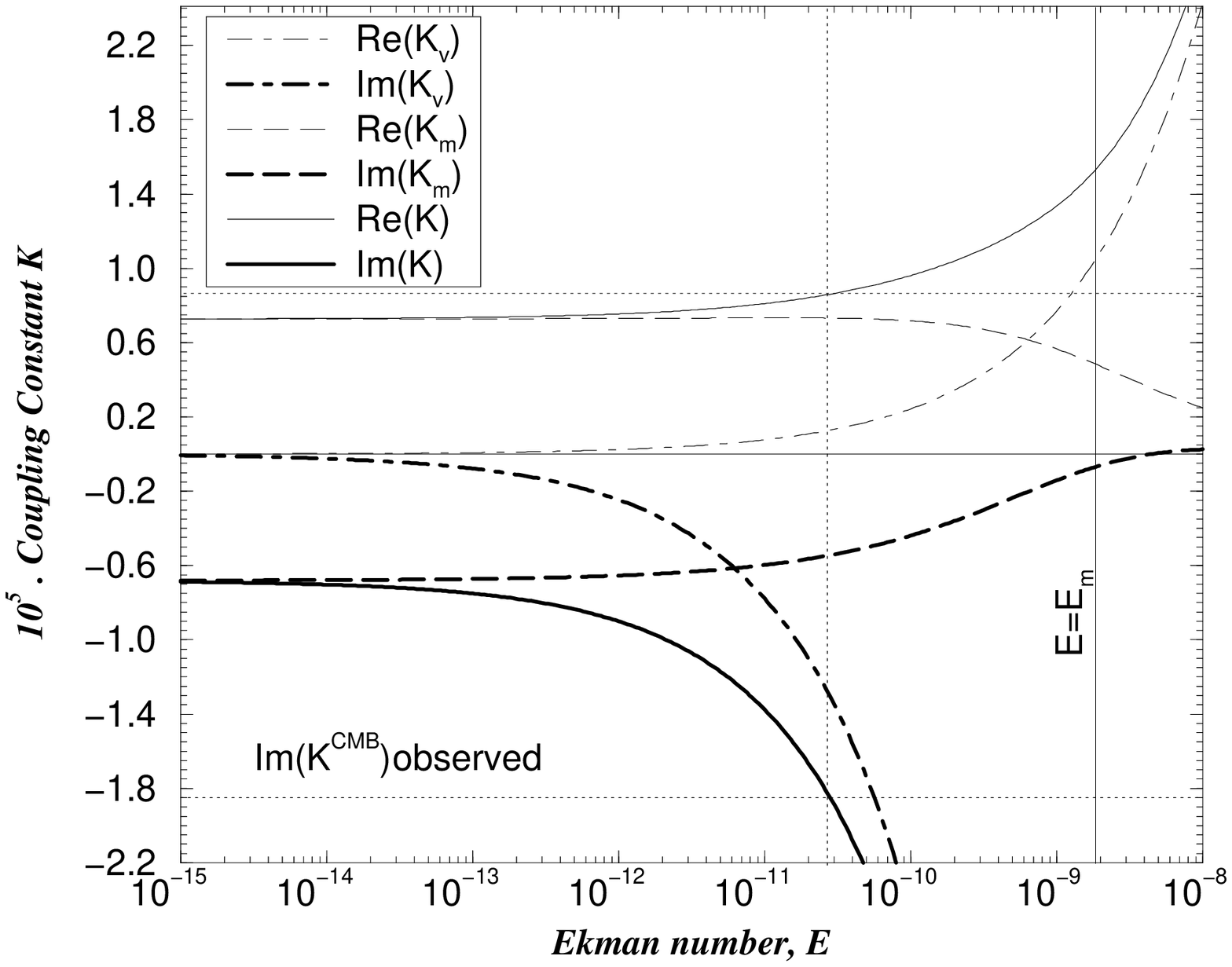}
	\end{center}
	\caption{Variations of the coupling constant $K$ as a function of the Ekman number. $K_v$ ($K_m$) is the viscous (magnetic) component of the coupling constant}
	\label{fig:varEk}
\end{figure}

This physical behavior is summarised on figure \ref{fig:varEk}. The magnetic torque does not vary for very low Ekman number ($E<10^{-11}$) and both components decrease as the Ekman number approaches the magnetic Ekman number. As expected, the viscous torque increases with the Ekman number. The imaginary part of the magnetic and viscous coupling constants become comparable for $E \approx 2.\;10^{-12}$ while their real parts match for a larger Ekman number ($E \approx 4.\;10^{-10}$). This difference results directly from the geometry of the spin over viscous torque which exhibits a very low imaginary part ($0.259$) compared to the real one ($2.62$) \cite{gree68}. As a conclusion, the resulting torque at the CMB is largely modified by a viscous shear layer for $E>10^{-12}$ even if the depth of the viscous layer is much smaller than the magnetic one (Figure \ref{fig:couche}a) and its effect barely changes the induced magnetic field at the boundary (Figure \ref{fig:couche}b).

In order to fit the imaginary part of the observed coupling constant ($-1.85 \;10^{-5}$), an Ekman number of 
$3.\;10^{-11}$ is necessary as shown on figure \ref{fig:varEk}. With such a value, the viscous dissipative process represents 85\% of the whole dissipation at the CMB but the real part of the coupling constant is still dominated (75\%) by the magnetic torque. 

\section{Electrical conductivity at the bottom of the mantle}

\begin{figure}
	\begin{center}
		\includegraphics [width = 12cm] {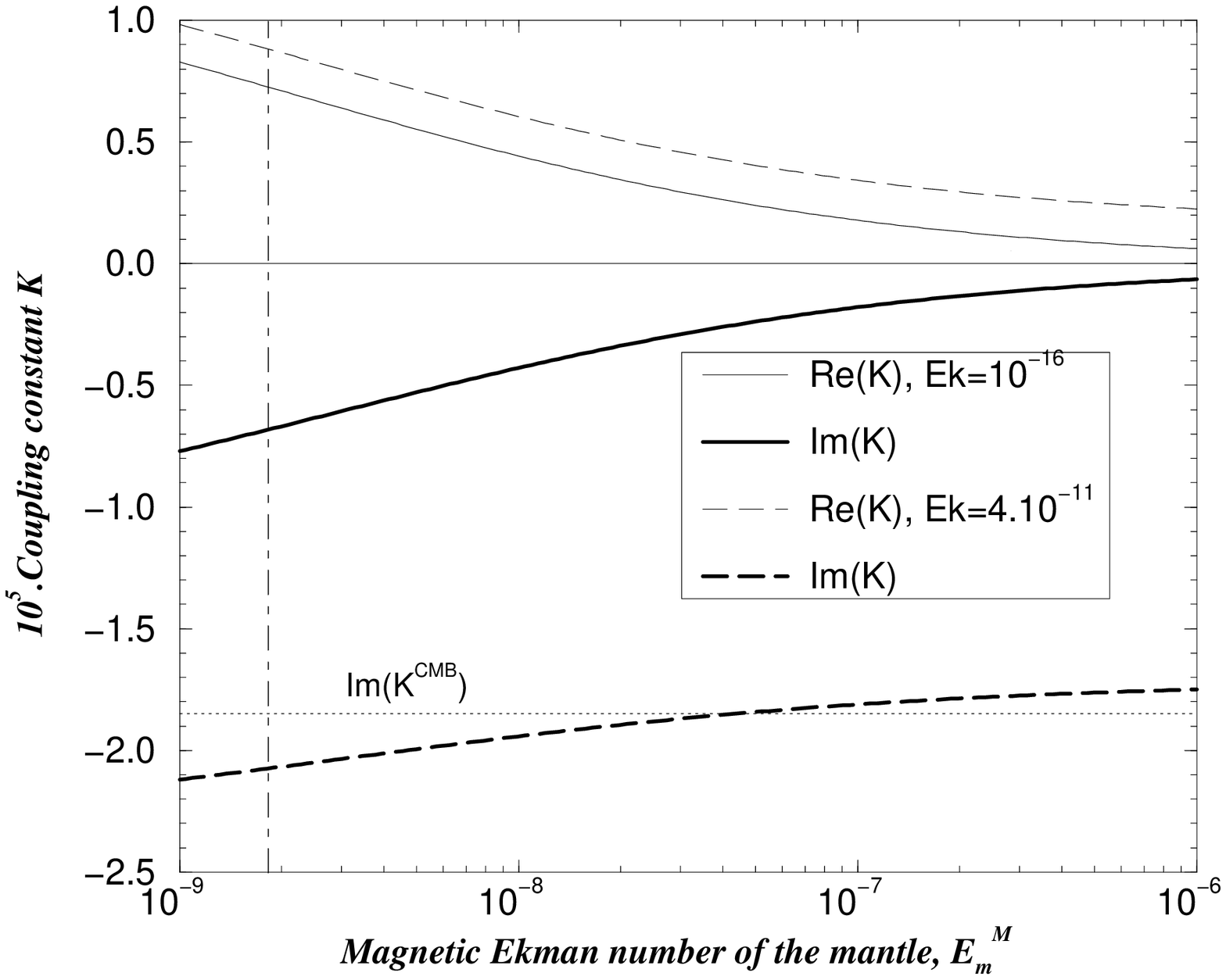}
	\end{center}
	\caption{Variations of the coupling constant with the Ekman magnetic number of the mantle.}
	\label{fig:varEm}
\end{figure}

In the visco-magnetic model for CMB parameters, the influence of the electrical conductivity of the lowermost mantle is secondary. Figure \ref{fig:varEm} shows the evolution of the coupling constant as the electrical resistivity is increased up to 10000 times the electrical resistivity of the core for an Ekman number of $4.\;10^{-11}$. The imaginary part of the coupling constant varies less than 20\% but the real part of the coupling constant is divided by 3. As expected,  for large magnetic Ekman number in the mantle, the torque is mainly dominated by the viscous part of the torque.

\begin{figure}
	\begin{center}
		\includegraphics [width = 12cm] {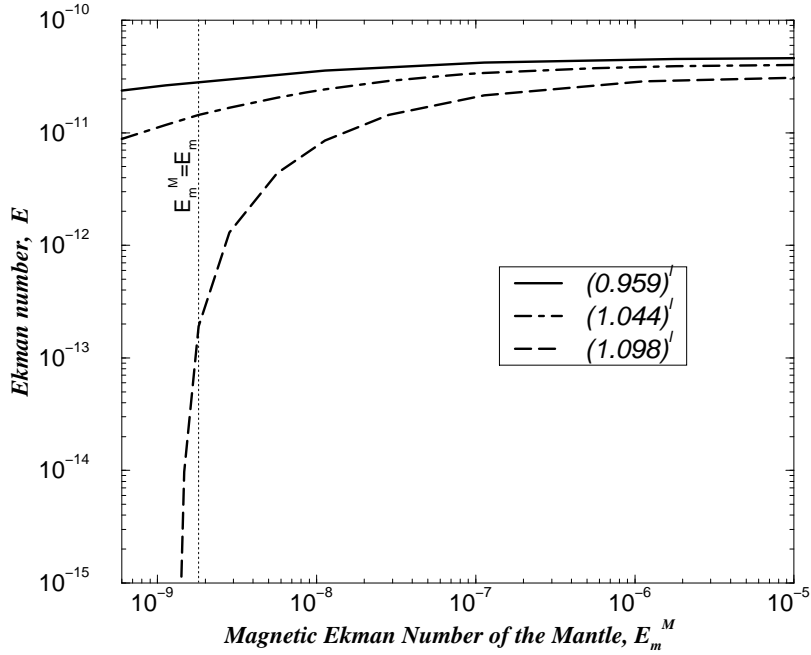}
	\end{center}
	\caption{Curves in the plan ($E,E_m^M$) on which $Im(K^{CMB}) = -1.85 10^{-5}$, for $E_m =1.8 \;10^{-9}$ and the three different spectra for ($l>13$). The observed magnetic field at the CMB corresponds to the nearly flat spectra $(0.959)^l$.}
	\label{fig:varEkEm}
\end{figure}

A trade off between viscous and magnetic torque could be found in order to fit the observational data of nutations. On figure \ref{fig:varEkEm}, for each value of the electrical conductivity of the mantle ($E_m^M$), we plot the Ekman number  ($E$) corresponding to a total torque in agreement with the observational constraint $Im(K^{CMB}) = -1.85 \; 10^{-5}$. For the nearly flat standard spectra $(0.959)^l$ (corresponding to the magnetic field at the CMB), Ekman numbers between $2$ and $5.\;10^{-11}$ are retrieved from the inversion whatever the conductivity at the bottom layer of the mantle. 
For the largest increasing spectra $(1.098)^l$, Ekman numbers vary more significantly with the conductivity of the lowermost mantle and very low Ekman numbers are retrieved when the electromagnetic torque becomes significant (comparable electrical conductivity on both sides of the CMB). 

The electrical conductivity at the bottom of the mantle is difficult to determine. Theoretical analysis and experimental measurements indicate that silicate rocks have a lower electrical conductivity than the liquid metal of the core \cite{poir92,shan93}. Discoveries of new phases of perovskite, such as post perovskite \cite{iita04}, or metal alloys of silicates may change this statement. The resistivity may even present lateral variations as shown by seismic lateral variations in the lowermost mantle \cite{lay98}. Except in the case of both a large electrical conductivity at the base of the mantle and a large magnitude of a hidden magnetic field at the CMB, the Ekman number needed to explain the observational data is around $10^{-11}$.
       
\section{Discussion}

The real part of the coupling constant is not dissipative and directly influences the period of the nutations \cite{deha97,hind00,math02}. But the discrepancy between the theoretical and observed periods of the free core nutation (FCN) is too large to be explained only by the real part of the coupling constant associated with the dissipative torque at the CMB. Hence, a dynamic ellipticity of a few hundred meters at the CMB has been introduced to account for this discrepancy. Then, the real part of the coupling constant cannot be used anymore as an observational constraint to determine the nature of the dissipative torque at the CMB. It is true, though, that a visco-magnetic dynamic model of the CMB reduces the real part of the coupling constant compared to a weak field model and consequently tends to increase by 10\% the estimate of the dynamic ellipticity at the CMB.

The visco-magnetic model of the magnetic skin layer shows that the small scales of the magnetic field at the CMB, that cannot be directly inferred from magnetic observations, do not contribute significantly to the electromagnetic torque between the core and the mantle. With a flat or decreasing power magnetic spectrum at the CMB, the electromagnetic torque is too weak to explain the coupling constant $Im(K^{CMB}) = -1.85 \;10^{-5}$, even if the electrical conductivity of the lower most mantle is comparable to the core one.

The visco-magnetic model of the boundary layers at the CMB using the observed geomagnetic field imposes the presence of viscous dissipative effects. Apparent Ekman numbers between $2$ and $4\;10^{-11}$ are needed to fit the observational constraint (corresponding to a turbulent viscosity of $3.5\;10^{-2} m^2s^{-1}$). This observational constraint may change as the quality of the data and their treatment improve \cite{flor00}. Only a reduction of the value of $Im(K^{CMB})$ by a factor 3 would make viscous effects unnecessary to explain the observations. However, the recent study of Palmer \& Smylie \shortcite{palm05}, if correct, gives an Ekman number of $7\;10^{-11}$ using their own analysis of VLBI data of the free core nutation and a pure viscous model of coupling at the CMB. The agreement is also very good with the results of Mathews \& Guo \shortcite{math05} which states that an Ekman number larger than $5\;10^{-11}$ is needed at the CMB. Such values of the Ekman number at the CMB are also compatible with the dissipation needed to account for the relaxation of torsional oscillations in the core \cite{zatm97,jaul03}.

At this stage, we like to see this observational constraint as a measurement of an apparent viscosity at the top of the core. The visco-magnetic model suggests an effective viscosity five thousand times larger than the expected molecular viscosity of iron at the core conditions. We would like to stress than an apparent viscosity is space and time dependent whereas, it is defined here on the whole surface of the CMB at the diurnal frequency. Consequently, this value of viscosity ($3.5\;10^{-2} m^2s^{-1}$) may not be generalized to the bulk of the core, at small scales and at different time scales. This is large but comparable to effective viscosities used in fluid dynamics of the ocean or the atmosphere. In these fields, a large apparent viscosity is the net result of the turbulent transport due to the small scales of the flow on the large scale flow. Such an explanation may be valid in the Earth's core even though we do not have any evidence for the action of small scales at the CMB. As discussed by Davies \& Whaler \shortcite{davi97}, the effective transport of momentum may be generated by convective motions associated to the dynamo process or by surfacic flows such as topographic winds, unstable boundary layers motions, or chemical/compositional fluxes.   

This work has been financed by the program DyETI and PNP of CNRS/INSU. The authors would like to thank Thierry Alboussi\`ere, Dominique Jault, Marianne Greff and V\'eronique Dehant and the rewievers for very helpful comments and suggestions.

\bibliography{couplage}
\bibliographystyle{gji}

\appendix
\section{Torque formulation}

The viscous torque  is computed from the viscous forces at the core mantle boundary: 
$$
 \Gamma_v =\int \!\! \int_{S}{\bf r} \times {\bf f}_v \, d{\bf S}
$$
where ${\bf f}_v$ is the viscous force per unit area. In our geometry and within the boundary layer approach, the viscous force on a sphere may be written : ${\bf f_v} = f_{v_{r\theta}} {\bf e_\theta} + f_{v_{r\varphi}}{\bf e_\varphi} = \frac{\p v_\theta}{\p r} {\bf e_\theta} + \frac{\p v_\varphi}{\p r} {\bf e_\varphi}$. At $r=1$, we have:
$$
{\bf r} \times {\bf f}_v=(-f_{v_{r\varphi}}\cos\theta\cos\varphi - f_{v_{r\theta}}\sin\varphi ){\bf e_x}+
(-f_{v_{r\varphi}}\cos\theta\sin\varphi + f_{v_{r\theta}}\cos\varphi){\bf e_y}+
f_{v_{r\varphi}}\sin\theta \;{\bf e_z}
$$

Introducing $\bar{\Gamma}=\Gamma_x + i\Gamma_y$, the dimensionless complex viscous torque at the CMB is:
$$
\bar{\Gamma}_v=\int_0^{\pi} \int_0^{2\pi}[i\frac{\p v_\theta}{\p r}-\cos\theta\frac{\p v_\varphi}{\p r}]\exp{(i\varphi)}\sin{\theta} d\theta d\varphi
$$
which in terms of $v_+$ and $v_-$ may be expressed:
$$
\bar{\Gamma}_v=\frac{i}{2}\int_0^{\pi} \int_0^{2\pi}[(1+\cos\theta) \frac{\p v_+}{\p r}+
(1-\cos\theta)
\frac{\p v_-}{\p r}]\exp{(i\varphi)}\sin{\theta}d\theta d\varphi
$$
which is the expression shown in equation (\ref{VT2}).
Equation (\ref{MT}) being similar to equation (\ref{VT}), the derivation of the magnetic torque is similar to that of the viscous torque.

\end{document}